\documentclass{aa}
\usepackage{graphicx}
\begin{document}
\title{Membership of 23 stars towards
the bulge globular clusters NGC 6528 and NGC 6553 
\thanks{Observations collected at the
European Southern Observatory - ESO, Chile}}

\author{
 P. Coelho\inst{1}
\and
 B. Barbuy\inst{1}
\and
 M.-N. Perrin\inst{2}
\and
 T. Idiart\inst{1}
\and
R. P. Schiavon\inst{3}
\and
S. Ortolani\inst{4}
\and
E. Bica \inst{5}
}
\offprints{P. Coelho}
\institute{
Universidade de S\~ao Paulo, Dept. de Astronomia,  CP 3386, 
S\~ao Paulo 01060-970, Brazil\\
 email: pcoelho@usp.br, barbuy@iagusp.usp.br, 
thais@iagusp.usp.br
\and
Observatoire de Paris, 61 av. de l'Observatoire, 75014 Paris, France\\
email: Marie-Noel.Perrin@obspm.fr
\and
Observat\'orio Nacional, rua General Jos\'e Cristino 77,
20921-400  Rio de Janeiro, Brazil and
Present address: UCO/Lick Observatory, University of California,
Santa Cruz CA 95064, USA\\
email: ripisc@ucolick.org
\and
Universit\`a di Padova, Dept. di Astronomia, Vicolo dell'Osservatorio 5,
I-35122 Padova, Italy\\
email: ortolani@pd.astro.it
\and
Universidade Federal do Rio Grande do Sul, Dept. de Astronomia,
CP 15051, Porto Alegre 91501-970, Brazil\\
email: bica@if.ufrgs.br
}

\date{Received 30 January 2001; accepted 11 June 2001}

\abstract{
Low resolution spectra of 23 stars towards the bulge globular
clusters NGC 6528 and NGC 6553 are analysed.
Radial velocities and atmospheric parameters are derived in order 
to check their membership in the clusters. Effective
temperatures were obtained from photometric data for stars with
T$_{\rm eff}$ $>$ 3800 K, whereas
for cooler stars, they were derived from
equivalent widths of TiO bands. Calibrations of
W(TiO) as a function of stellar parameters based on 
a grid of synthetic spectra are presented.
Metallicities were derived from a comparison
of the observed spectra to a grid of synthetic spectra.
The sample comprises evolutionary stages from the 
Red Giant Branch to the Horizontal Branch, with parameters in the range
3200 $\leq$ $T_{\rm eff}$ $\leq$ 5000 K and -0.5 $\leq$ log g $\leq$ 2.4.
The mean metallicities obtained for NGC 6528 and NGC 6553 are 
[Fe/H] $\approx$  -0.5 and -0.7, in both cases with [Mg/Fe] = +0.3;
assuming the same overabundance for the $\alpha$ elements
 O, Mg, Si, S, Ca and Ti, 
this gives [Z/Z$_{\rm \odot}$] = -0.25 and -0.45.
Membership verification by means of low resolution spectra is a 
crucial step in preparing targets for high resolution spectroscopy
with 8m class telescopes.
}

\titlerunning{Membership towards
NGC 6528 and NGC 6553}
\authorrunning{P. Coelho et al.}
\maketitle


\section{Introduction}

The study of the stellar populations in the Galactic bulge is very
important to constrain possible models of galaxy formation. In particular,
the determination of the metallicities and abundance ratios of bulge
stars, either from the field or in clusters, provides key information to
help decide among the possible scenarios for the history of chemical
enrichment of the Galaxy.  

There are few previous studies of radial velocities and
metallicity estimations of bulge stars from low resolution
spectra. In the analysis of 400 field bulge stars by 
Sadler et al. (1996), metallicities and [Mg/Fe] values
were estimated. Minniti (1995a,b) studied the membership
of stars towards 7 bulge globular clusters.
	 
The best studied among the bulge globular clusters are NGC
6528 ($\alpha_{2000}$ = 18$^{\rm h}$04$^{\rm m}$49.6$^{\rm s}$,
$\delta_{2000}$ = -30$^{\rm o}$03'20.8") and NGC 6553 ($\alpha_{2000}$
= 18$^{\rm h}$09$^{\rm m}$15.7$^{\rm s}$, $\delta_{2000}$ = -25$^{\rm
o}$54'27.9"). Ortolani et al. (1995) have shown that, besides being old,
these clusters have luminosity functions which are very similar to 
that of Baade's Window, which indicates that they belong to the same
stellar population.

NGC 6528 is located in the Baade Window, at a distance d$_{\odot}$ =
7.83 kpc from the Sun, and NGC 6553 is relatively close to the Sun, at
a distance  d$_{\odot}$ =  5.1 kpc (Barbuy et al. 1998). As they are
both located in crowded fields, the measurement of radial velocities of
individual stars is of crucial importance for the determination of
their membership in the clusters.

Both clusters are known to be metal-rich. However, there is no consensus
in the literature regarding their detailed metal abundances. Recently,
Barbuy et al. (1999) analysed high resolution spectra of two giant
stars of NGC 6553. An Iron abundance of [Fe/H] = -0.55$\pm$0.2 and
abundance ratios [Na/Fe] $\approx$ [Al/Fe] $\approx$ [Ti/Fe] $\approx$
+0.5, [Mg/Fe] $\approx$ [Si/Fe] $\approx$ [Ca/Fe] $\approx$ +0.3
 were derived. These ratios imply an overall
metallicity [Z/Z$_{\rm \odot}$] $\approx$ -0.1.
Cohen et al. (1999), analysing high resolution spectra of five red
horizontal branch stars, obtained a mean metallicity [Fe/H] = -0.16 and
an excess of the $\alpha$-element calcium to iron of about 0.3 dex, which
imply an overall metallicity [Z/Z$_{\rm \odot}$] $\approx$ +0.1. 
Metal abundances of these clusters are also discussed
in Barbuy et al. (1999) and Barbuy (2000).

In view of the disagreement between previous determinations
of [Fe/H], it is important that abundance estimations be extended to
a larger number of stars of both clusters. In this paper, we determine
radial velocities, effective temperatures, gravities and
estimations  of metallicities [Fe/H] based on low resolution
spectra for 23 stars towards NGC 6553 and NGC 6528, and verify
their membership in these clusters.

In Sect. 2 the observations are described.
The radial velocities derived are presented in Sect. 3.
In Sect. 4
the stellar parameters are derived, and synthetic spectra
are compared to observations to estimate metallicities. 
Concluding remarks are given in Sect. 5.

\section{Observations}

Low resolution spectra of individual stars
of NGC 6528 and NGC 6553 were obtained
in 1992 August and 1994 June, at 
the 1.5m ESO telescope at ESO (La Silla). 
The Boller \& Chivens spectrograph was employed.
In 1992 August the Thompson CCD \# 18 with 1024$\times$1024
pixels, with a pixel size of 19 $\mu$m was used. 
A resolution of $\Delta\lambda$ $\sim$ 8 {\rm \AA} and a spectral coverage
of $\lambda\lambda$ 4800-8800 {\rm \AA} were achieved.
In 1994 June, the Ford Aerospace FA 2048 L, frontside illuminated, uncoated
CCD detector (ESO \# 24) with 2048$\times$2048 pixels and pixel size  
15$\times$15 $\mu$m was used. The grating \# 27 resulted in a
spectral resolution $\Delta\lambda$ $\sim$ 4 {\rm \AA}
and a spectral coverage in the range $\lambda\lambda$ 
4800-7550 {\rm \AA}.

The log of observations is provided in Table 1.
The stars are identified according to the charts by
Hartwick (1975) for NGC 6553 and van den Bergh \&
Younger (1979) for NGC 6528. Spectra of a given star 
were co-added by weighting their S/N ratios; the 
final S/N  are indicated in Table 1.

\begin{table}
\caption[1]{Log of observations}
\begin{flushleft}
\begin{tabular}{lccccc}
\noalign{\smallskip}
\hline
\noalign{\smallskip}
Star & V & V-I & Exp. (s) & Date & S/N  \\
\noalign{\smallskip}
\hline
\noalign{\smallskip}
\noalign{NGC 6528}
\noalign{\smallskip}
I 1 & 16.10 & 1.93 & 5400 & 06.08.92 & 190 \\
I 2 & 15.73 &2.59  & 1800 & 16.06.94 & 25\\
    &       &      & 5400 & 06.08.92 & 75 \\
I 5 & 15.37 &2.22 &  2700 & 17.06.94 & 20  \\
I 6 & 15.89 &3.54 &  2700 & 17.06.94 & 40  \\
II 8 & 15.71 &2.19 &  2100 & 17.06.94 & 20  \\
II 14 & 15.76 &3.47 &  2700 & 17.06.94 & 10  \\
I 23 & 17.19 &1.70 &  1800 & 16.06.94 & 10  \\
I 24 & 16.89 &1.66 &  4500 & 17.06.94 & 20  \\
I 25 & 16.11 &2.09 &  4500 & 16,17.06.94 & 40  \\
I 27 & 15.90 &3.08 &  1800 & 16.06.94 & 20 \\
     &        &    &  5400 & 06.08.92 & 100 \\
I 36 & 16.41 &1.98 &  2100 & 17.06.94 & 20  \\
II 39 & 15.88 &2.30 &  2100 & 17.06.94 & 30  \\
I 40 & 15.93 &2.08 &  2100 & 17.06.94 & 20  \\
I 42 & 16.42 &2.15 &  2100 & 17.06.94 & 30 \\
II 70 & 15.85 &2.36 &  1800 & 16.06.94 & 20  \\
\noalign{\smallskip}
\hline
\noalign{\smallskip}
\noalign{NGC 6553}
\noalign{\smallskip}
III 2 & 16.89 &1.95 & 1800 & 16.06.94 & 20  \\
III 3 & 15.82 &2.41 &  3600 & 14,16.06.94 & 25 \\
      &       & &  5400 & 07.08.92    & 100 \\
III 17& 15.36 &3.01 &  1800 & 14.06.94 & 30  \\
II 51 & 15.48 &2.54 &  1800 & 17.06.94 & 10 \\
      &       & &  5400 & 07.08.92 & 80 \\
II 52 & 16.84 &1.93 &  1800 & 17.06.94 & 40\\
      &       &     & 5400  &  07.08.92 & 230 \\
II 85 & 15.52 &2.51 &  3600 & 14,16.06.94 & 55  \\
II 94 & 15.44 &3.38 &  1800 & 17.06.94 & 20  \\
II 95 & 15.73 &2.64 &  1800 & 17.06.94 & 35  \\
\noalign{\smallskip} 
\hline \end{tabular}
\\
\end{flushleft} 
\end{table}

In Figs. 1 and 2 are shown the $V$ vs. $V-I$ Colour-Magnitude Diagrams 
of NGC 6528 and NGC 6553 using data obtained with the Hubble Space 
Telescope (Ortolani et al. 1995) where the sample stars are identified.

\begin{figure}[]
\centering\includegraphics[width=8.2cm]{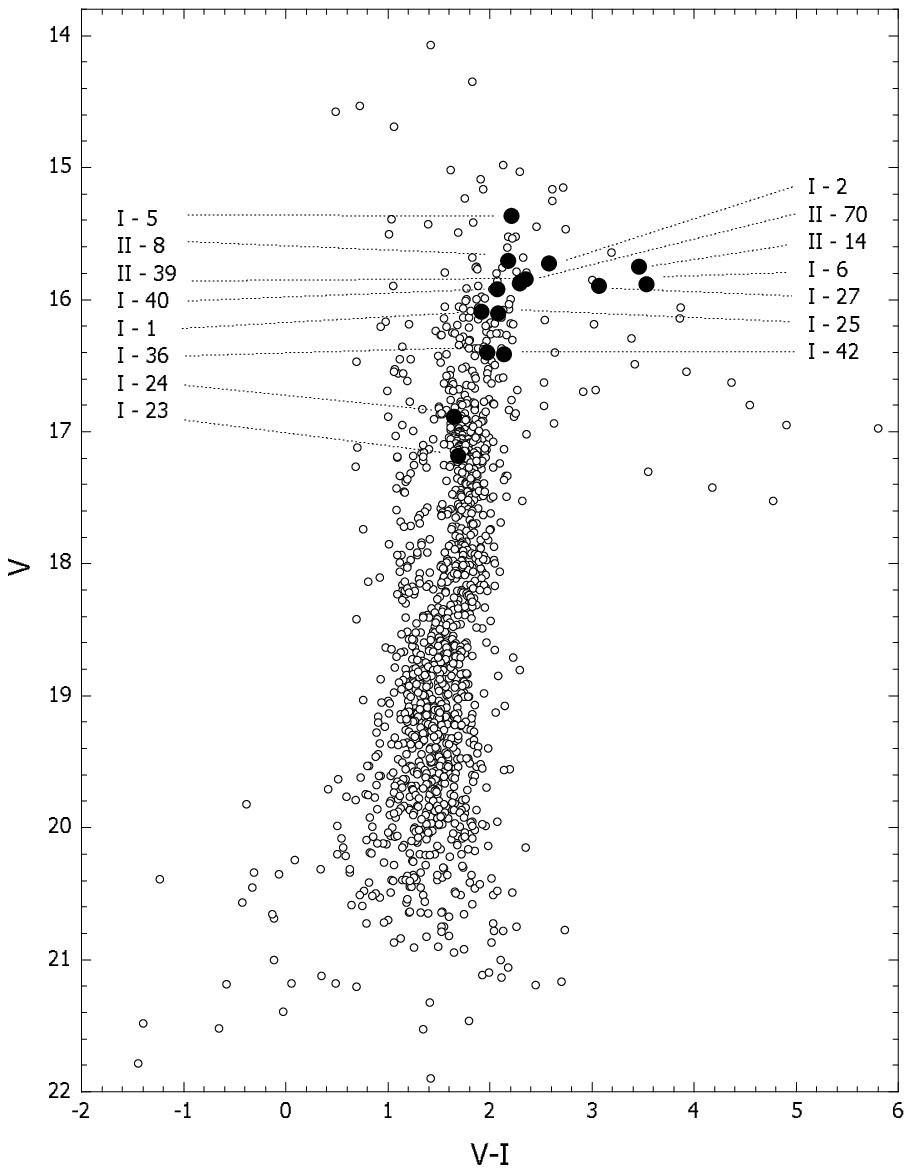}
\centering\includegraphics[width=8.5cm]{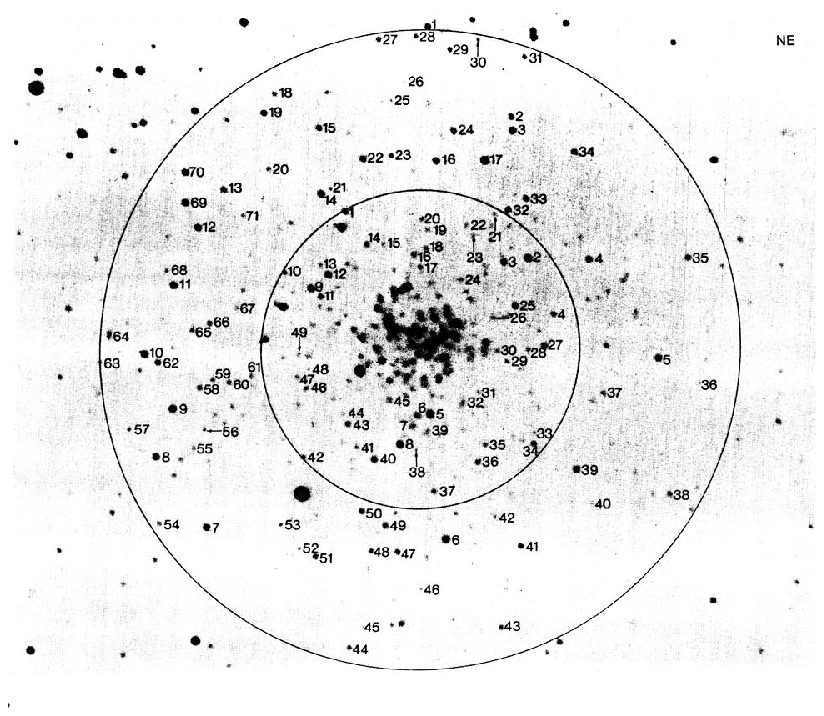}
\caption[ ]{
{\it Top}: HST Colour-magnitude diagram of NGC 6528, where the observed
stars are indicated as filled circles; {\it Bottom}:
A map of the cluster scanned from van den Bergh \& Younger (1979).}
\label{n6528 cmd}
\end{figure}

\begin{figure}[]
\centering\includegraphics[width=8.2cm]{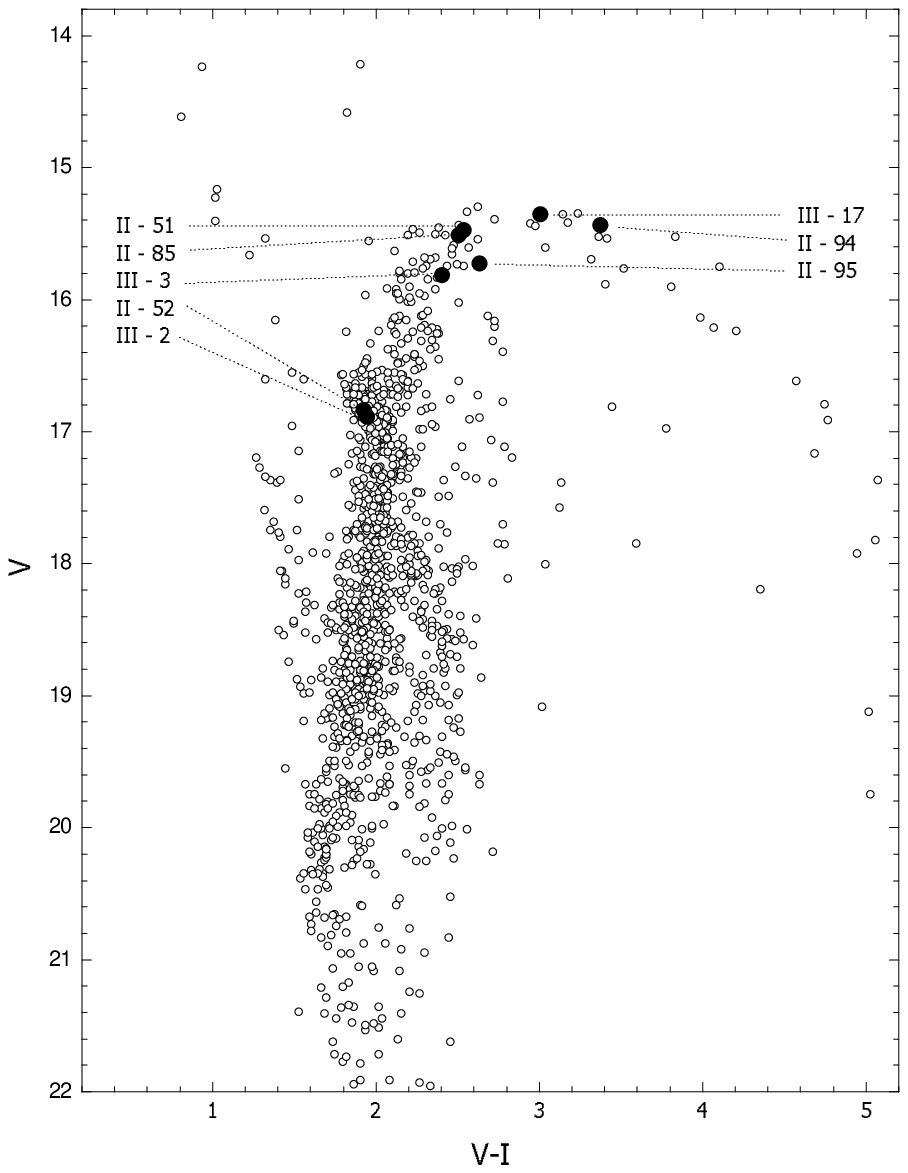}
\centering\includegraphics[width=8.2cm]{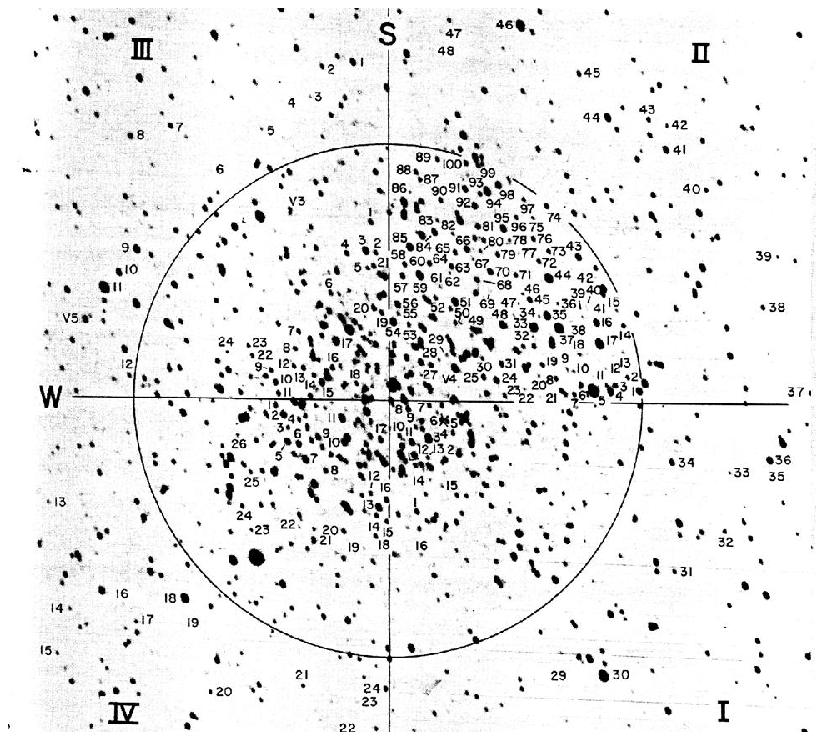}
\caption[ ]{
{\it Top}: HST Colour-magnitude diagram of NGC 6553, where the observed
stars are indicated as filled circles; {\it Bottom}:
A map of the cluster scanned from Hartwick (1975). 
}
\label{n6553 cmd}
\end{figure}


\section{Radial velocities}

The radial velocities were determined by means of three methods, 
as explained below and reported in Table 2. 
The observed radial velocities derived  
were transformed to heliocentric
values using the observation dates given in Table 1.

(a) A Fourier cross-correlation was applied on the program spectra 
relative to selected template spectra.
As templates, 12 G, K and M stars were selected from the Jacoby et al.'s (1984)
library, which have approximately the same spectral resolution 
(4.5 {\rm \AA}) of the sample spectra.
The templates adopted were the ones closest in spectral type
to each of the program stars. 
The spectra of both sample and template stars were normalized
and different regions in the spectra were
defined in order to give highest peaks of cross-correlation for each 
considered template. 
The results obtained with this
method are given in column 2 of Table 2.
The r.m.s. of the values derived with
each template spectrum is of the order of 15 km s$^{-1}$.
A systematic effect was identified 
for the coolest stars,
since all of them appeared to show lower
velocities when compared to the hotter stars of the same cluster
and these values were not considered.

(b) Mean shifts between the observed 
wavelengths of identified absorption lines and laboratory wavelengths
were measured (column 3 of Table 2). The r.m.s. 
of the values derived 
is of the order of 15 km s$^{-1}$.

(c) the code HALO (Cayrel et al. 1991) 
derives radial velocities
by comparing the observed spectrum to a grid of synthetic spectra,
using a cross-correlation technique. The grid of synthetic spectra
available (Barbuy et al. 2001) does not contain stars cooler than 
T$_{\rm eff}$ $<$ 4000 K, and for this reason the errors should
be higher for velocities of stars
cooler than T$_{\rm eff}$ $<$ 3700 K in which TiO bands are
pronounced. 


\begin{table}
\caption[1]{Heliocentric 
radial velocities v$_{\rm r}$ (km s$^{-1}$) obtained with
different methods: (a) cross-correlation relative to observed
template spectra using IRAF; (b) mean shift in wavelength
for a list of selected lines; (c) cross-correlation relative to
synthetic spectra using the code HALO (see text).}
\begin{flushleft}
\begin{tabular}{lccc}
\noalign{\smallskip}
\hline
\noalign{\smallskip}
Star \hspace* {13 pt}&  \multispan3 \hfill  v$_{\rm r}$ (km s$^{-1}$) \hfill\\
     &   method (a) \hspace* {13 pt} & method (b)  \hspace* {13 pt}  & method (c) \\
\noalign{\smallskip}
\hline
\noalign{\smallskip}
\noalign{NGC 6528} 
\noalign{\smallskip}
I 1  &  --- & 262  &  174 \\
I 2  &  224  &  262   & 208 \\
I 5  &  ---  &  261  & 224 \\
I 6  &  232   &  246   & 225 \\
II 8 &  289  &  283   & 271 \\
II 14&  ---  &  264  & 227 \\
I 23 &  221 &  251   & ---\\
I 24 &  238  &  220   & 202 \\
I 25 &  265  &  257   & 231 \\
I 27 &  --- &  237   & 188 \\
I 36 &  244  &  249   & 229 \\
II 39&  30  &  17  &  7 \\
I 40 &  246 &  236 & 244 \\
I 42 &  197 &  235 &  212 \\
II 70&  230 &  263 &  --- \\
\noalign{\smallskip}
\hline
\noalign{\smallskip}
\noalign{NGC 6553}
\noalign{\smallskip}
III 2 & -2  & -7  & -25 \\
III 3 & 5  & 17  & 18 \\
III 17& 3  & 8  & -19 \\
II 51 & ---  & 60  & --- \\
II 52 & -16  &-28  & -33 \\
II 85 & 35  & 56  & -4 \\
II 94 &-42  &-56  & -63 \\
II 95 & 11  &-12  & -9 \\
\noalign{\smallskip}
\hline
\end{tabular}
\end{flushleft}
\end{table}

\begin{figure}[]
\includegraphics[width=8.2cm]{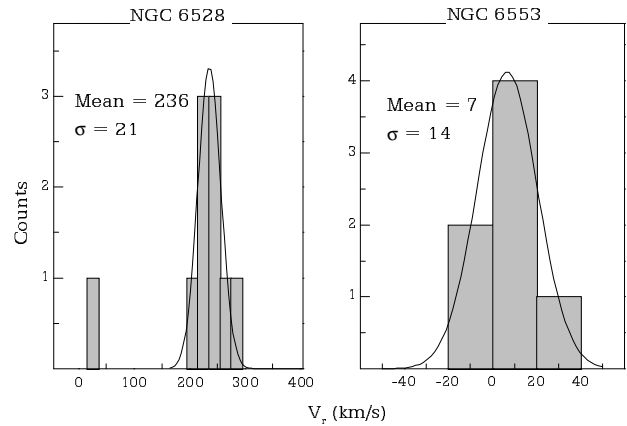}
\caption[]{Histograms of radial velocities obtained for the stars
with the cross-correlation technique using IRAF, where  
the gaussian fits are presented. The deviant point in the histogram
of NGC 6528 is the star II 39, which is probably a non-member.}
\label{hist vr}
\end{figure}

Histograms of radial velocities of individual stars (coolest stars excluded)
corresponding to each method were built.
Gaussian curves were fitted to each histogram, from which the radial 
velocity corresponding to each method was derived, as 
reported in Table 3
together with values from the literature.
An example of this procedure is presented in Fig. \ref{hist vr} for the 
cross-correlation technique using IRAF. 
The final radial velocities adopted for the clusters correspond 
to the mean of the values
derived from the three methods.

\begin{table}
\caption[]{Radial velocities of NGC 6528 and NGC 6553 given in the 
literature and values derived in the present work, for which 
the standard deviation values obtained with the gaussian fits are 
indicated in parentheses.}
\label{vrlit}
\begin{tabular}{ccc}
\hline
\noalign{\smallskip}
\multispan2 \hspace* {36 pt} v$_{\rm r}$ ( km s$^{-1}$ ) \hspace* {36 pt} & \hspace* {11 pt} Reference \hspace* {11 pt} \\
\hspace* {12 pt} NGC 6528 \hspace* {12 pt}& \hspace* {12 pt} NGC 6553 \hspace* {12 pt}&   \\  
\noalign{\smallskip}
\hline
218   & 6.4 & 1,2 \\ 
164.8 & -24.5 & 3 \\ 
208   & 48 & 4 \\
212   & 8.4 & 5 \\
189 & --- & 6 \\
160   & -5 & 7 \\
143 & -12 & 8 \\
236 (21) & 7 (14)& 9 \\
248 (11) & 1 (16) & 10 \\
217 (9) & -10 (13) & 11 \\
\noalign{\smallskip}
\hline
\end{tabular}
\smallskip
\\
\footnotesize References to Table: 1 Barbuy et al. (1999); 2 Barbuy (2000);
3 Harris (1996); 4 Minniti (1995) (mean values excluding non-member
stars); 5 Rutledge et al. (1997);
6 Armandroff \& Zinn (1988); 7 Zinn \& West (1984);
8 Zinn (1985); 9 present paper (cross-correlation method with
IRAF; 10 present paper (mean wavelength shift with IRAF);
11 present paper (code HALO by a cross-correlation
method)
\end{table}


\section{Stellar parameters}

\subsection{Temperatures}

The effective temperatures were estimated
from $B-V$, $V-I$, $V-K$ and $J-K$ colours,
based on the colour vs. T$_{\rm eff}$
calibrations by Bessell et al. (1998),
which in turn are based on NMARCS models by Plez et al. (1992) and their grid
extensions. These effective temperatures are listed in Table 5.
V and I colours were obtained with the Hubble Space Telescope
(Ortolani et al. 1995) and  J and K colours were obtained with 
the detector IRAC2 at the 2.2m telescope of
ESO (Guarnieri et al. 1998).

For NGC 6528 the colour excesses adopted were E(V$-$I) = 0.68 and
E(B$-$V) = 0.52 (Barbuy et al. 1998). For NGC 6553 
E(V$-$I) = 0.95 and E(B$-$V) = 0.7 were adopted (Guarnieri et al. 1998).
The $V-K$ and $J-K$ colours were dereddened assuming E(V-K)/E(B-V) = 2.744 
and E(J$-$K)/E(B$-$V)=0.527 (Rieke \& Lebofsky 1985). 

\begin{figure}[]
\includegraphics[width=8.2cm]{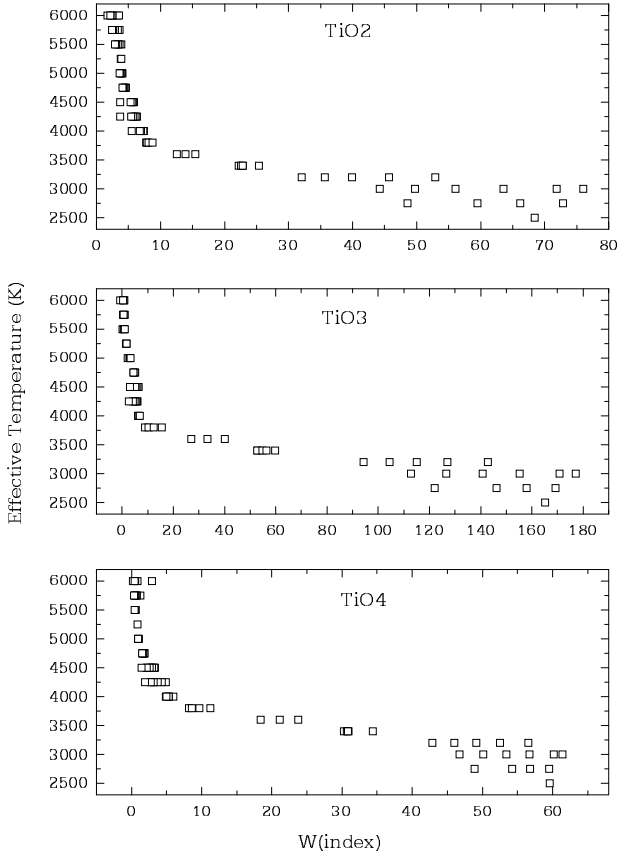}
\caption[ ]{
TiO indices measured on the synthetic spectra as a function of effective 
temperatures. 
These measurements correspond to spectra convolved with FWHM = 8 
${\rm \AA}$}
\label{calib TiO}
\end{figure}

\begin{figure}[]
\includegraphics[width=8.2cm]{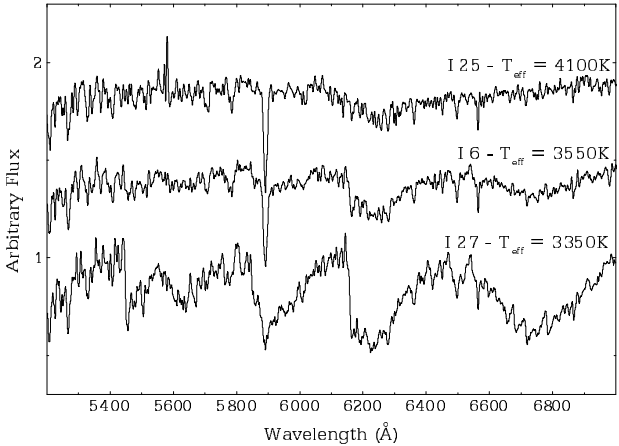}
\caption[ ]{
Spectra of three individual stars of NGC 6528 with different temperatures. 
It is clear that the T$_{\rm eff}$ = 3350 K star has more pronounced TiO
molecular bands with respect to the T$_{\rm eff}$ = 4100 K one.}
\label{diftem}
\end{figure}

An independent method for the derivation of temperatures was based 
on calibrations of equivalent widths of TiO bands. The indices as defined
in Table \ref{inTiO} were measured on the grid of synthetic spectra 
by Schiavon \& Barbuy (1999) in the range of parameters
3000 $\leq T_{\rm eff} \leq$ 5000 K, 
$-$0.5 $\leq$ log g $\leq$ 2.5 and [Fe/H] = 
$-$0.3. These indices are shown in Fig. \ref{calib TiO} 
for a resolution of $\Delta\lambda$ = 8 {\rm \AA}. 
Polynomial curves of the form
T$_{\rm eff}$ = f(W(TiO)) were derived and applied 
to the indices measured in the sample stars.
The TiO indices are strongly sensitive to temperature for 
$T_{\rm eff} \leq$ 3800 K
as illustrated in Fig. \ref{diftem}. For $T_{\rm eff} \geq$ 4000 K a degeneracy appears
due to the fact that TiO bands are not present at these higher
temperatures.

A more general polynomial of the form 
log W(TiO) = 
({a + b log T$_{\rm eff}$ + c log $g$ + d [Fe/H] 
+ e (log T$_{\rm eff}$)$^{2}$
+ f [Fe/H]$^{2}$ + g [Fe/H] log T$_{\rm eff}$}),
valid in the range 2500  $\leq$   T$_{\rm eff}$ $\leq$ 5000 K,
-0.5 $\leq$ log g $\leq$ 2.5 and -0.5 $\leq$ Fe/H $\leq$ 0
was derived.
The coefficients of the formula above are shown in Table \ref{coeffs} for 
convolutions of $\Delta\lambda$(FWHM) = 4 {\rm \AA} and 8 {\rm \AA}.

\begin{table}
\caption[]{TiO indices used in the derivation of 
effective temperatures. The definition of these indices 
is such as 
to avoid regions of strong telluric lines.}
\label{inTiO}
\begin{flushleft}
\begin{tabular}{lccc}
\hline
\noalign{\smallskip}
Index & Blue continuum & Bandpass & Red continuum \\
\noalign{\smallskip}
\hline
\noalign{\smallskip}
TiO2 & 6033.6-6050.6 & 6300.0-6455.0 & 6525.0-6538.0 \\
TiO3 & 6525.0-6539.0 & 6617.6-6860.0 & 7036.0-7046.6 \\
TiO4 & 7036.0-7046.6 & 7053.0-7163.0 & 7534.2-7546.8 \\
\noalign{\smallskip}
\hline
\end{tabular}
\end{flushleft}
\end{table}

\begin{table}[]
\caption[]{Effective temperatures  
based on calibrations of $B-V$, $V-I$, $V-K$ and $J-K$ colours 
(columns 2 to 5) and TiO indices (columns 6 to 8), and
final values adopted.
The discrepancy between the photometric 
and spectroscopic effective temperatures of II-51 might be explained
by observations in different phases of the light curve of this
probable red variable.}
\label{temps}
\begin{flushleft}
\begin{tabular}{l@{}c@{}c@{}c@{}c@{}c@{}c@{}c@{}c@{}}
\hline
\noalign{\smallskip}
 & \multicolumn{8}{c}{Effective Temperatures (K)}\\
Star \hspace*{6 pt} & \small $V-I$ \hspace*{1.5 pt}& \small $B-V$ \hspace*{1.5 pt}& \small $J-K$ 
\hspace*{1.5 pt}& \small $V-K$ \hspace*{1.5 pt}& TiO2 \hspace*{1.5 pt}& TiO3 
 \hspace*{1.5 pt}& TiO4 \hspace*{1.5 pt}& Final\\
\noalign{\smallskip}
\hline
\noalign{\smallskip}
\multicolumn {2}{l}{NGC 6528}\\
\noalign{\smallskip}
I 1    & 4305 & 4449 & ---     & ---     & --- &  ---   & --- & 4400 \\
I 2    & 3673 & 3883 & ---     & ---     & 3922 & 3698 & 3506 & 3700 \\
I 5    & --- & --- & ---     & ---     & 3468 & 3053 &  ---   & 3250 \\
I 6    & 3451 & 3890 & ---     & ---     & 3612 & 3585 & 3405 & 3550 \\
II 8   & 3963 & 3992 & ---     & ---     & 3776 & 4049 & 3789 & 3950 \\
II 14  & --- & --- & ---     & ---     & 3468 & 3053 &  ---   & 3250 \\
I 23   & 4783 & 4770 & ---     & ---     & --- &  ---   & --- & 4800 \\
I 24   & 4883 & 4707 & ---     & ---     & --- &  ---   & --- & 4800 \\
I 25   & 4074 & 4092 & ---     & ---     & --- &  ---   & --- & 4100 \\
I 27   & --- & --- & ---     & ---     & 3673 &  ---   & 3001 & 3350 \\
I 36   & 4218 & 4216 & ---     & ---     & --- &  ---   & --- & 4200 \\
I 40   & 4084 & 4137 & ---     & ---     & --- &  ---   & --- & 4100 \\
I 42   & 3999 & 4149 & ---     & ---     & --- &  ---   & --- & 4050 \\
II 70  & 3810 & 3951 & ---     & ---     & --- & --- &  ---   & 3900 \\
\noalign{\smallskip}
\hline
\noalign{\smallskip}
\multicolumn {2}{l}{NGC 6553}\\
\noalign{\smallskip}
III 2  & 4828 & ---     & 4809 & 4559 & --- & 3842 & 3776 & 3800 \\
III 3  & 4015 & ---     & 4221 & 3939 & 3782 & 3800 & 3775 & 3800 \\
III 17 & 3606 & ---     & 3966 & 3611 & 4052 & 3698 & 3503 & 3750 \\
II 51  & 3885 & ---     & 3966 & 3856 & 3281 & 3206 &  ---   & 3250 \\
II 52  & --- & ---     & --- & --- & 4241 & 3703 & 3752 & 3900 \\
II 85  & 3912 & ---     & 4140 & 3835 & --- &  ---   & --- & 3950 \\
II 94  & 3506 &  ---    & 3984 & 3432 & 4802 & 4408 & 3832 & 3650 \\
II 95  & --- &  ---    & --- & --- &  ---   &  ---   &  ---   & 6500 \\
\noalign{\smallskip}
\hline
\end{tabular}
\end{flushleft}
\end{table}

The temperatures obtained 
and final values adopted are reported in Table 5. 
Photometric temperatures were adopted for stars for which
the TiO temperature T$_{\rm TiO}$  $\geq$ 3800 K, 
whereas for stars with T$_{\rm TiO}$ 
$<$ 3800 K  the mean of TiO temperatures were adopted.

\begin{table}
\caption{Coefficients of polynomial fits of W(TiO) indices
as a function of stellar parameters for 
2500 $\leq$ T$_{\rm eff}$ $\leq$ 5000 K, 
-0.5 $\leq$ log g $\leq$ 2.5 and -0.5 $\leq$ [Fe/H] $\leq$ 0}
\label{coeffs}
\begin{tabular}{l@{}r@{\hspace{6 pt}}r@{\hspace{6 pt}}r@{\hspace{6 pt}}r@{\hspace{6 pt}}r@{\hspace{6 pt}}r@{\hspace{6 pt}}}
\hline
\noalign{\smallskip}
Coefficient & \multispan2 TiO2 &
\multispan2  TiO3  & \multispan2  TiO4 \\
$\Delta\lambda$ & 4{\rm \AA}  &  8{\rm \AA} 
&  4{\rm \AA} &  8{\rm \AA} &  4{\rm \AA} &  8{\rm \AA}  \\

\noalign{\smallskip}
\hline
\noalign{\smallskip}
a (constant)                  & 81.49  & 81.89  & 55.70  & 56.51  & -175.11  & -158.38 \\
b (log(T$_{\rm eff}$))        & -39.44 & -39.71 & -22.93 & -23.53 & 105.33   & 95.92   \\
c (log $g$)                   & -0.04  & -0.04  & -0.04  & -0.05  & -0.06    & -0.06   \\
d ([Fe/H])                    & 9.28   & 9.14   & 3.40   & 3.21   & -7.42    & -5.10   \\
e (log(T$_{\rm eff}$)$^2$)    & 4.75   & 4.79   & 2.17   & 2.27   & -15.66   & -14.34  \\
f ([Fe/H]$^2$)                & -0.47  & -0.45  & -0.40  & -0.38  & -0.45    & -0.08   \\
g ([Fe/H]$\times$                     &        &        &        &        & \\
\phantom{----}log(T$_{\rm eff}$)) & -2.56  & -2.51  & -0.87  & -0.82  & 2.16    & 1.54    \\
\hline
\noalign{\smallskip}	
$\chi$$^2$                    &  0.92   & 0.92   & 0.90   & 0.90   & 0.97    & 0.96    \\	
\noalign{\smallskip}
\hline
\end{tabular}
\end{table}

\subsection{Gravities}

Gravities were derived using the  classical relation 
log g$_*$=  4.44 + 4log T$_*$/T$_{\odot}$ + 
0.4(M$_{\rm bol}$-M$_{\rm bol \odot}$)
+ log M$_*$/M$_{\odot}$, 
adopting T$_{\odot}$ = 5770 K, M$_*$ = 
0.8 M$_{\odot}$ and M$_{\rm bol \odot}$ = 4.74 cf. Bessell et al. (1998).
For deriving M$_{\rm bol *}$ we used the distance modulus adopting
a total extinction A$_{\rm V}$ =
2.43 for NGC 6553 and A$_{\rm V}$ = 1.8 for NGC 6528 (Barbuy et al. 1998).
The bolometric magnitude corrections were taken from Bessell et al. (1998).

The resulting M$_{\rm bol *}$ and gravities are given in Table 7.
 Taking into consideration the errors due to uncertainties 
in T$_{\rm eff}$ and M$_{bol}$
the final error in log g is estimated to be of $\pm$0.5 dex.

\subsection{Metallicities}

Spectrum synthesis calculations 
were used to fit the observed spectra.
The calculations of synthetic spectra were carried out using
the code described in Barbuy et al. (2000)
where molecular lines of MgH A$^2$$\Pi$-X$^2$$\Sigma$, 
CH A$^2$$\Delta$-X$^2$$\Pi$,
CN A$^2$$\Pi$-X$^2$$\Sigma$,
C$_2$ Swan A$^3$$\Pi$-X$^3$$\Pi$  and TiO
$\alpha$ C$^3$$\Delta$-X$^3$$\Delta$,  
$\gamma$ A$^3$$\Phi$-X$^3$$\Delta$ and
$\gamma$' B$^3$$\Pi$-X$^3$$\Delta$
systems are taken into account.

For atomic lines the laboratory oscillator strengths by Fuhr et
al. (1988), Martin et al. (1988), Wiese et al. (1969),
and laboratory values compiled by McWilliam \& Rich (1994) were adopted
whenever available, otherwise they were taken from 
fits to the solar spectrum  (see discussion in Barbuy et al.
1999).

ATLAS9 and NMARCS models were employed. A grid of models using
the ATLAS9 code (Kur\'ucz 1993) was created adopting a mixing
length parameter $\alpha$ = 0.5
(see Barbuy et al. 2001).
NMARCS photospheric models for giants by Plez et al. (1992)
and their unpublished extended grids were employed (see 
more details in Schiavon \& Barbuy 1999).

The metallicities were obtained based on two methods,
both using synthetic spectra:

(i) The observed spectra were compared to synthetic spectra in
the range $\lambda\lambda$ 5000--7500
 ${\rm \AA}$. 
The metallicities were estimated by interpolating between 
synthetic spectra of [Fe/H]= 0.0, --0.3, --0.5 and --0.6,
in all cases assuming [Mg/Fe] = +0.3,
and the temperatures and gravities determined in Sects.
4.1 and 4.2.

\begin{figure}[]
\includegraphics[width=8.2cm]{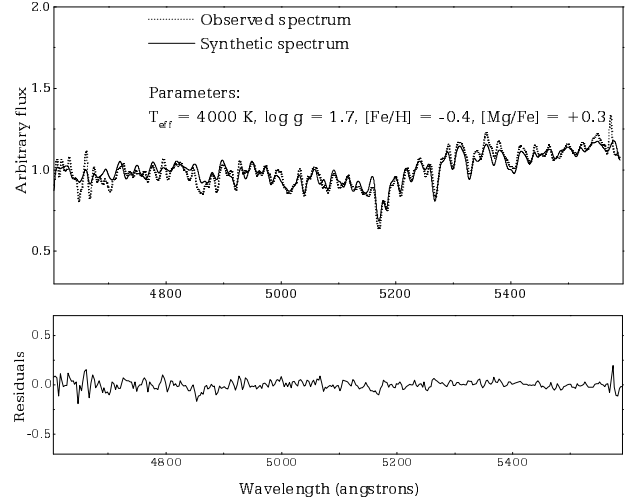}
\caption[ ]{
Analysis of NGC 6528 I-1 employing the code HALO: {\it Top:} observed
spectrum (dotted line) and synthetic spectrum (solid line)
computed with  T$_{\rm eff}$ = 4400 K,
 log $g$ = 1.7, [Fe/H] = -0.4 and [Mg/Fe] = +0.30
 (FWHM = 8 {\rm \AA}); {\it Bottom:} Residual flux relative to
the template synthetic spectrum. }
\label{bx6528-1}
\end{figure}

(ii) Comparisons with a grid of synthetic spectra
in the wavelength region $\lambda\lambda$ 4600-5600
{\rm \AA}, using the differences method as described in Cayrel et al.
(1991) and Barbuy et al. (2001), are carried out.
In this method, the observed spectrum is divided by 
a reference synthetic spectrum.  The resulting signal
can be expressed as a linear combination of variations in 
temperature, gravity and metallicity.
In conjunction with the grid of synthetic spectra, it is possible to establish
the differences in T$_{\rm eff}$, log g and [Fe/H] 
between the program star and the reference synthetic spectrum
through a perturbation method.
The grid covers the range 4000 $\leq$ T$_{\rm eff}$
$\leq$ 7000 K, 0.0 $\leq$ log g $\leq$ 5.0, 
-3.0 $\leq$ [Fe/H] $\leq$ +0.3,  and [Mg/Fe] = 0.0 and
+0.4. 
Fig. \ref{bx6528-1} shows the fit to NGC 6528 I-1.

In Table 7 are listed the temperatures, gravities,
[Fe/H] and [Mg/Fe] obtained with methods (i) and (ii).
Note that method (ii) tends to give lower metallicities relative to
method (i). This may be due to limitations of the grid of synthetic
spectra, which is being extended to cover wider ranges of parameters.

The stars NGC 6528 I-5,
NGC 6553 III-2 and II-95 are probable non-members,
given that their atmospheric
parameters are incompatible with their location in the 
Colour-Magnitude Diagrams of the clusters.
The star II-51 appears to be  too cool (Table 7) with respect
to its CMD locus (Fig. 2). However, considering that it 
could be a red variable, this star is tentatively
classified as a possible member.

Mean metallicities of [Fe/H] = -0.5 for NGC 6528 and
[Fe/H] = -0.7 for NGC 6553 were estimated from
gaussian fits to the histograms of metallicities given
in Table 7 (typical standard deviations are of 0.2 dex).
These metallicities, together with the magnesium excess of
 [Mg/Fe] $\approx$ +0.3, and assuming that all other $\alpha$
elements show an excess of +0.3 dex relative to iron,
 result in overall metallicities
of [Z/Z$_{\rm \odot}$] = -0.25 and -0.45 for NGC 6528 and NGC 6553
respectively. 

\begin{table}
\caption[]{Bolometric magnitudes, atmospheric parameters
and method employed (see Sect. 4.3).
M$_{\rm bol}$ values were not reported for probable non-member stars.
}
\label{paramspaula}
\begin{flushleft}
\begin{tabular}{lccccccccc}
\hline
\noalign{\smallskip}
Star & M$_{\rm bol}$ & $T_{\rm eff} $ & log g & [Fe/H] &[Mg/Fe] & Method \\
\noalign{\smallskip}
\hline
\noalign{\smallskip}
\multicolumn {2}{l}{NGC 6528}\\
\noalign{\smallskip}
I 1  & -0.8  & 4400 & 1.7 & -0.4 & 0.3 & i\\
     &       & 4400 & 1.7 & -0.4 & 0.3 & ii\\
I 2  & -2.0  & 3700 & 0.9 & -0.5 & 0.3 & i\\
I 5  & --    & 3250 & 0.7 & -0.5 & 0.3 & i\\
I 6  & -3.0  & 3550 & 0.4 & -0.6 & 0.3 & i\\
II 8 & -1.5  & 4000 & 1.2 & -0.5 & 0.3 & i\\
     &       & 4000 & 0.5 & -0.6 & 0.3 & ii\\
II 14 &-3.1  & 3250 & 0.2 & -0.5 & 0.3 & i\\
I 23 & 0.6   & 4800 & 2.4 & 0    & 0.3 & i\\
     &       & 4800 & 2.4 & -0.4 & 0.2 & ii\\
I 24 & 0.4   & 4800 & 2.3 & 0    & 0.3 & i\\
     &       & 4800 & 2.3 & -0.3 & 0.1 & ii\\
I 25 & -1.0  & 4100 & 1.4 & -0.6 & 0.3 & i\\
     &       & 4100 & 1.4 & -1.1 & 0.2 & ii\\
I 27 & -2.5  & 3350 & 0.5 & -0.3 & 0.3 & i\\
I 36 & -0.6  & 4200 & 1.7 & -0.6 & 0.3 & i\\
     &       & 4250 & 1.0 & -0.8 & 0.4 & ii\\
I 40 & -1.2  & 4100 & 1.4 & -0.4 & 0.3 & i\\
     &       & 4100 & 1.4 & -0.7 & 0.3 & ii\\
I 42 & -0.8  & 4050 & 1.5 & -0.4 & 0.3 & i\\
     &       & 4050 & 1.5 & -1.2 & 0.2 & ii\\
II 70 & -1.6 & 3900 & 1.1 & -0.6 & 0.3 & i\\
\noalign{\smallskip}
\hline
\noalign{\smallskip}
\multicolumn {2}{l}{NGC 6553}\\
\noalign{\smallskip}
III 2 & --     & 3800 & 2.0 & -0.7 & 0.3 & i\\
III 3 & -1.1   & 3800 & 1.3 & -0.7 & 0.3 &  i\\
III 17& -2.3 &   3750 & 0.8 & -0.7 & 0.3 &  i\\
II 51 & --   & 3250 & 0.8 & -0.4 & 0.3  &  i\\
II 52 & 0.6   & 3900 & 2.0 & -0.6 & 0.3 &  i\\
II 85 & -1.5   & 3950 & 1.2 & -0.6 & 0.3 & i\\
II 94 & -2.7   & 3650 & 0.6 & -1.1 & 0.3  &  i\\
II 95 & --  & 6500 & 2.7 & -0.4: & ---    &  ii\\
\noalign{\smallskip} 
\hline 
\end{tabular}
\end{flushleft} 
\end{table}

\section{Conclusions}

The study of individual stars in globular clusters along
their evolutionary stages is of prime importance for an
improved understanding of stellar evolution. Low resolution
spectroscopy provides a means for the study of a large number
of stars. In the present work we have measured radial
velocities and estimated metallicities
in 23 stars towards the globular clusters NGC 6528
and NGC 6553,
which allows us to identify member stars.
We also obtained their atmospheric properties to a first 
approximation.
This is an important step before applying efforts to obtain 
high resolution spectroscopy with 8m class telescopes.
The method presented here is also of interest for
last generation multi-object instruments such as
VLT-VIMOS.

The stars were analysed by comparisons between their
observed spectra and a grid of synthetic spectra.
TiO equivalent widths were used to estimate effective temperatures of
stars  cooler  than $T_{\rm eff}$ $\leq$ 3800 K and a calibration of
 equivalent widths of TiO bands as a function of atmospheric parameters
 is  presented.

Mean values of heliocentric radial velocities of 
v$_{\rm r}$ = 234 km s$^{-1}$ for NGC 6528 and
v$_{\rm r}$ = -1 km s$^{-1}$ for NGC 6553 are derived.

Regarding membership, among the 23 stars observed we concluded that
4 of them are probable non-members. These are: NGC 6528 II-39, non-member due to
a deviant radial velocity, and NGC 6528 I-5, NGC 6553 III-2 and II-95, non-members due to
incompatibilities of atmospheric parameters vs. 
location in the Colour-Magnitude Diagrams.

NGC 6553 II-51 could be a non-member, or a red variable for
which the spectrum was taken during a cool phase.

The basic stellar parameters derived show the interesting
result that 
there is a trend for
member giants of NGC 6528 
to be more metal-poor
than the two Horizontal Branch stars NGC 6528 I-23 and I-24,
thus reproducing the discrepancy found between analysis
of NGC 6553 giants by Barbuy et al. (1999) and Horizontal
Branch stars by Cohen et al. (1999). 
Given the errors involved in the analysis of low resolution spectra,
these results have to be checked with high resolution spectra, and
further studies of
this discrepancy will be possible only with a homogeneous
analysis of stars ranging from
the red giant branch to the HB, and also employing 
different sets of model atmospheres all along the evolutionary
sequence.

In summary, we obtained for NGC 6528 and NGC 6553 metallicities
of [Fe/H] = -0.5 $\pm$ 0.3 and
[Fe/H] = -0.7 $\pm$ 0.3. Using [Mg/Fe] $\approx$ +0.3,
and assuming that other $\alpha$ elements show the same excess of 
+0.3 dex relative to iron, the results are
[Z/Z$_{\rm \odot}$] = -0.25 and -0.45 for NGC 6528 and NGC 6553
respectively.

\begin{acknowledgements}
We are grateful to A. Milone for having 
carried out part of the observations.
We acknowledge partial financial support from CNPq
and Fapesp. P. Coelho and T. Idiart acknowledge
respectively the Fapesp Master
fellowship n$^{\rm o}$ 98/07492-4,
and  Post-Doc fellowship n$^{\rm o}$ 97/13083-7.
RPS acknowledges support provided by
the National Science Foundation through grant GF-1002-99 and from the 
Association of Universities for
Research in Astronomy, Inc., under NSF cooperative agreement AST 96-1361.
\end{acknowledgements}


%
%


\begin{thebibliography}{}
\bibitem[1988]{armandroffzinn} Armandroff T.E., Zinn R., 1988, AJ 96, 92
\bibitem[1998a]{barbuy98a} Barbuy B., Bica E., Ortolani S., 1998, A\&A 333, 117
\bibitem[1999]{barbuy00} Barbuy B., 2000, in {\it The Chemical
Evolution of the Milky Way: Stars vs. Clusters}, Eds. F.
Matteucci, Kluwer Acad. Pub., in press
\bibitem[1999]{barbuy99} Barbuy B., Renzini A.,  Ortolani S., Bica E.,
Guarnieri M.D.,  1999, A\&A 341, 539
\bibitem[2001]{barbuy01} Barbuy B., Perrin M.-N., Katz D.,
      Cayrel R., Spite M., van 't Veer-Menneret C., 2001, A\&A, 
      submitted
\bibitem[]{} Bessell M.S., Castelli F., Plez B., 1998, A\&A 337, 321
\bibitem[]{}Carretta E., Gratton R.G., 1997, A\&AS 121, 95
\bibitem[]{}Cohen J.G., Gratton R.G., Behr B.B., Carretta E., 1999,
ApJ 523, 739
\bibitem{} Fuhr J.R., Martin G.A., Wiese W.L., 1988,
{\it Atomic Transition Probabilities: Iron through Nickel},
Journal of Physical and Chemical Reference Data, vol. 17,  no. 4
\bibitem[1998]{guarnieri} Guarnieri M.D., Ortolani S., Montegriffo P., 
Renzini A., Barbuy B., Bica E., Moneti A., 1998, A\&A 331, 70
\bibitem[]{} Cayrel R., Perrin M.-N., Barbuy B., Buser R., 1991, A\&A 247, 108
\bibitem {harris} Harris W.E., 1996,  AJ 112, 1487
\bibitem[]{}Hartwick F.D.A., 1975, PASP 87, 77
\bibitem{jacoby} Jacoby G.H., Hunter D.A., Christian C.A., 1984, ApJ 419,
592
\bibitem{kurucz} Kur\'ucz, R., 1993, CD-ROM 18
\bibitem{} Martin G.A., Fuhr J.R., Wiese W.L., 1988,
{\it Atomic Transition Probabilities: Scandium through Manganese},
Journal of Physical and Chemical Reference Data, vol. 17, no. 3
\bibitem{} McWilliam A., Rich R.M., 1994, ApJS 91, 749 (MR)
\bibitem[1995]{minniti95a} Minniti D., 1995a, A\&A 303, 468
\bibitem[1995]{minniti95b} Minniti D., 1995b, A\&AS 113, 299
\bibitem[1999]{sch99} Schiavon R.P., Barbuy B., 1999, ApJ 510, 934
\bibitem[1995b]{ortolani95} Ortolani S., Renzini A., Gilmozzi R., Marconi 
G., Barbuy B., Bica E., Rich R.M., 1995, Nature 377, 701
\bibitem{} Plez B., Brett J.M., Nordlund ${\rm \AA}$, 1992, A\&A 256, 551
\bibitem{} Rieke G.H., Lebofsky M.J., 1985, ApJ 288, 618
\bibitem{} Rutledge G.A., Hesser J.E., Stetson P.B., 
Mateo M., Simard L., Bolte M., Friel E.D., Copin Y.,
 1997,  PASP 109, 883
\bibitem[1996]{sadler} Sadler E., Rich R.M., Terndrup D.M., 1996, AJ 112, 171 
\bibitem[1979]{vdb} van den Bergh S., Younger F., 1979, AJ 84, 1305
\bibitem{} Wiese W.L., Martin G.A., Fuhr J.R.,  1969,
{\it Atomic Transition Probabilities: Sodium through Calcium},
           NSRDS-NBS 22
\bibitem[1985]{zinn} Zinn R., 1985, ApJ 293, 424
\bibitem[1984]{zinn84} Zinn R., West, M.J., 1984, ApJS 55, 45
\end{thebibliography}
\end{document}